\documentclass[11pt,a4paper]{article}

\usepackage[margin=1in]{geometry}
\usepackage{graphicx}
\usepackage{xcolor}
\usepackage{tocloft}
\usepackage{enumitem}
\usepackage{setspace}
\usepackage{titlesec}
\usepackage{amssymb}
\usepackage[unicode=true]{hyperref}
\usepackage[superscript,sort,compress]{cite}

\hypersetup{
    colorlinks=true,
    linkcolor=blue,
    citecolor=black,
    bookmarksopen=true,
    bookmarksnumbered=true
}

\titleformat{\section}{\large\bfseries}{}{0em}{}
\titleformat{\subsection}{\bfseries}{}{0em}{}

\usepackage{caption}
\begin{document}
{\Large \textbf{Regulating oxygen content and superconductivity in La$_3$Ni$_2$O$_{7+\delta}$}}\\[1em]
\textbf{Peiyue Ma,\textsuperscript{1,13} Jingyuan Li,\textsuperscript{1,13} Xing Huang,\textsuperscript{1,2,13} Yixing Zhao,\textsuperscript{3} Yifeng Han,\textsuperscript{3} Mengwu Huo,\textsuperscript{1} Deyuan Hu,\textsuperscript{1} Chaoxin Huang,\textsuperscript{1} Hengyuan Zhang,\textsuperscript{1}, Sihao Deng,\textsuperscript{4} Lunhua He,\textsuperscript{4} Juan Rodriguez-Carvajal,\textsuperscript{5} Abhisek Bandyopadhyay,\textsuperscript{6,7} Alessandro Puri,\textsuperscript{8,9} Devashibhai Adroja,\textsuperscript{7,10} Xiang Chen,\textsuperscript{1} Tao Xie,\textsuperscript{1} Zhen Chen,\textsuperscript{2,11} Hualei Sun.\textsuperscript{12,*} Meng Wang\textsuperscript{1,*} }

\textsuperscript{1}Guangdong Provincial Key Laboratory of Magnetoelectric Physics and Devices, Institute of Neutron Science and Technology,  School of Physics, Sun Yat-Sen University, Guangzhou 510275, Guangdong, China\\
\textsuperscript{2}Beijing National Laboratory for Condensed Matter Physics, Institute of Physics, Chinese Academy of Sciences, Beijing 100190, China\\
\textsuperscript{3}School of Chemistry and Chemical Engineering, Hainan University, Haikou 570228, China\\
\textsuperscript{4}Spallation Neutron Source Science Center, Dongguan 523803, Guangdong, China\\
\textsuperscript{5}Diffraction Group, Institut Laue-Langevin, 71, Avenue des Martyrs, 20156-38042 Grenoble Cedex 9, France\\
\textsuperscript{6}Department of Physics, Ramashray Baleshwar College (A Constituent Unit of Lalit Narayan Mithila University, Darbhanga), Dalsingsarai, Samastipur, Bihar 848114, India\\
\textsuperscript{7}ISIS Neutron and Muon Facility, STFC, Rutherford Appleton Laboratory, Didcot, Oxfordshire, OX11 0QX, United Kingdom\\
\textsuperscript{8}Department of Physics and Astronomy, Alma Mater Studiorumâ Universita di Bologna, Viale Berti Pichat 6/2, 40127 Bologna, Italy\\
\textsuperscript{9}CNR - Istituto Officina dei Materiali, Grenoble c/o ESRF, The European Synchrotron, 71 Avenue des Martyrs, 40220-38043 Grenoble Cedex 9, France\\
\textsuperscript{10}Highly Correlated Matter Research Group, Physics Department, University of Johannesburg, Auckland Park 2006, South Africa\\
\textsuperscript{11}School of Physical Sciences, University of Chinese Academy of Sciences, Beijing 100049, China\\
\textsuperscript{12}School of Science, Sun Yat-Sen University, Shenzhen 518107, Guangdong, China\\
\textsuperscript{13}These authors contributed equally\\
\textsuperscript{*}Correspondence: wangmeng5@mail.sysu.edu.cn; sunhlei@mail.sysu.edu.cn

\textbf{GRAPHICAL ABSTRACT}

\begin{figure}[h]
    \centering
    \includegraphics[width=0.85\linewidth]{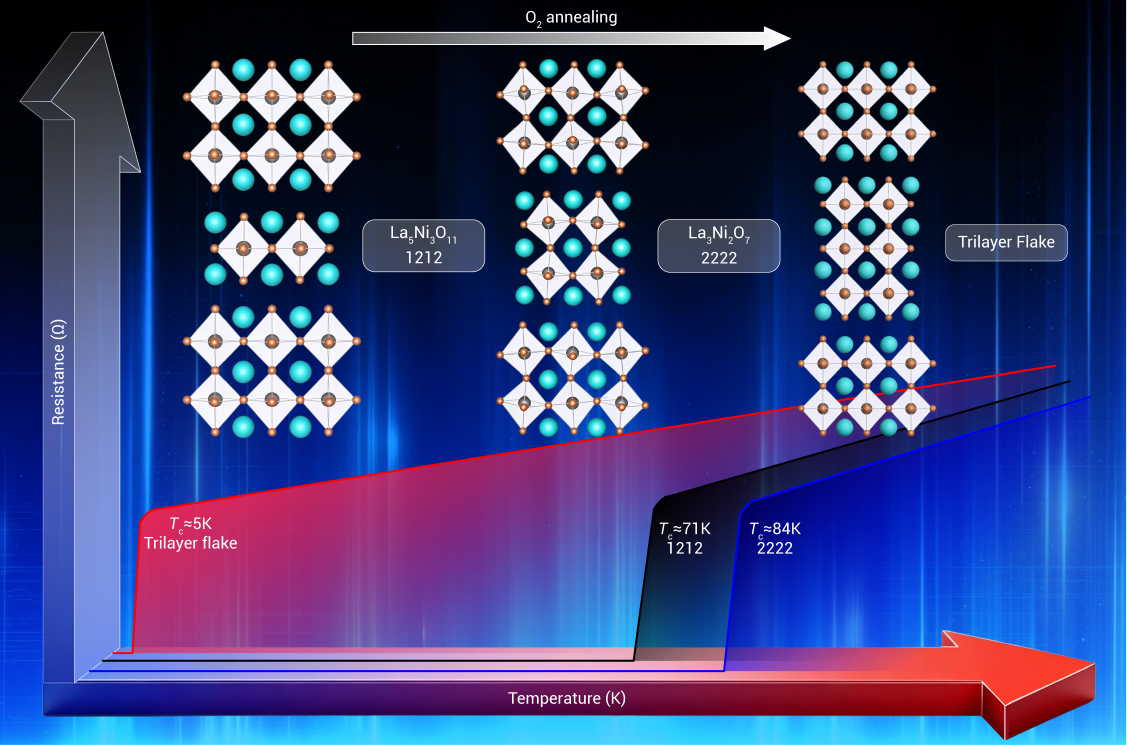}
    \label{fig:graphical-abstract}
\end{figure}

\textbf{PUBLIC SUMMARY}

\begin{itemize}[leftmargin=*, label=$\blacksquare$]
\item High‑quality Ruddlesden‑Popper nickelates advance our understanding of their superconductivity.

\item Synthesizing high‑quality nickelates is challenging due to oxygen inhomogeneity and coexisting mixed phases.

\item This work advances La$_3$Ni$_2$O$_{7+\delta}$ synthesis and achieves controllable oxygen modulation.

\item Oxygen stoichiometry and microstructure affect superconductivity and upper critical field in nickelates.

\item We built a phase diagram of superconducting temperature and upper critical field versus oxygen content.

\end{itemize}

\textbf{ABSTRACT}\\
The synthesis of high-quality Ruddlesden-Popper (RP) nickelates remains challenging due to variations in oxygen content and the prevalence of intergrown RP phases. Precisely controlling the stoichiometry and characterizing the resulting physical properties are essential for understanding the mechanism of high-$T_c$ superconductivity in these materials. In this work, we synthesize a series of La$_3$Ni$_2$O$_{7+\delta}$ samples with systematically controlled averaged oxygen content across the range $-0.34 \le \delta \le 0.08$ and perform comprehensive structural and compositional analyses. Tuning the averaged oxygen content drives the evolution of the microstructure from a mixture of bilayer and monolayer-bilayer phases, to a pure bilayer phase, and finally to a predominantly bilayer phase containing trilayer intergrowths. High-pressure transport measurements reveal distinct superconducting transitions with contrasting $T_c$ values, corresponding to the bilayer phase, the hybrid phase, and trilayer inclusions. Notably, we find that the phase purity also modulates the upper critical field ($H_{c2}$) of the bilayer superconductivity. By establishing a phase diagram of $T_c$ and $H_{c2}$ as functions of averaged oxygen content in La$_3$Ni$_2$O$_{7+\delta}$, this work advances synthetic control and provides new insights into the superconducting mechanism of RP nickelates.

\vspace{1em}


\vspace{1em}

\textbf{SUBJECT AREA:}\\
Condenced Matter Physics

\textbf{INTRODUCTION}

After the discovery of high-temperature superconductivity in cuprates, nickelates emerged as promising candidates for analogous superconductivity owing to similarities in their crystal and electronic structures. \cite{Anisimov1999,Wang2024nor} It was not until 2019 that superconductivity was observed in Nd$_{0.8}$Sr$_{0.2}$NiO$_2$ thin films, exhibiting a transition temperature of 15 K.\cite{Li2019} This material features Ni$^{1.2+}$ ions with spin configuration and electronic structure analogous to doped cuprates with Cu$^{2+}$ ions.\cite{Anisimov1999,GU2022su} A subsequent breakthrough was the discovery of 80 K superconductivity in the bilayer Ruddlesden-Popper (RP) nickelate La$_3$Ni$_2$O$_7$ under pressure, which contains Ni$^{2.5+}$ ions.\cite{Sun2023,Zhang2024h,Wang2024p,Qiu2026} Following that, superconductivity has been observed in pressurized trilayer La$_4$Ni$_3$O$_{10}$,\cite{Zhu2024s,Li2024str,Zhang2025sc} the hybrid monolayer-bilayer (1212) phase La$_5$Ni$_3$O$_{11}$,\cite{Li2024de,Shi2025pr} the hybrid monolayer-trilayer (1313) phase La$_3$Ni$_2$O$_7$,\cite{Wang2024lo,Chen2024p,Huang2025s} and corresponding thin film samples at ambient pressure.\cite{Ko2025,Zhou2025a,Hao2025,Wang2026p} The RP nickelates, denoted by the formula La$_{n+1}$Ni$_n$O$_{3n+1}$ for $n=1, 2, 3, ..., \infty$, have thus emerged as a new family of high $T_c$ superconductors.\cite{Zhang1994,Wang2024nor,Sakurai2026,Zhou2025gi,Nie2026sup} In contrast to the well-controlled phase purity and stoichiometry of cuprates and iron-based superconductors, the intergrowth of RP phases, variations in oxygen content, and the pressure-dependent nature of superconductivity in RP nickelates pose significant challenges to understanding the underlying mechanisms.\cite{Dong2024,Puphal2026sup,Wang2025recent}

Density functional theory suggests that metallization of the $\sigma$-bond in Ni-3$d_{z^2}$, which hybridizes with the interlayer apical oxygen 2$p$ orbitals, is intimately linked to the emergence of superconductivity under pressure\cite{Sun2023,Luo2023b}. Synchrotron X-ray diffraction (SXRD) reveals a structural transition in which the out-of-plane Ni-O-Ni bond angle increases from 168$^\circ$ to 180$^\circ$, coinciding with the onset of superconductivity and highlighting the importance of interlayer exchange interactions\cite{Wang2024b,Li2025i}. Resonant inelastic X-ray scattering and inelastic neutron scattering experiments confirm a dominant interlayer exchange interaction in bilayer La$_3$Ni$_2$O$_7$\cite{Chen2024ele,Xie2024s}. In this scenario, removing the interlayer apical oxygen atoms would disrupt the Ni-O-Ni superexchange pathway and thus suppress superconductivity\cite{Wang2026sup}. Scanning transmission electron microscopy (STEM) studies show that the interlayer apical oxygen atoms in La$_3$Ni$_2$O$_7$ are susceptible to loss, resulting in oxygen vacancy defects\cite{Dong2024}. The appearance of pressure-induced superconductivity and its associated superconducting volume fraction are directly influenced by these defects, providing further evidence that apical oxygen atoms are critical for superconductivity. Moreover, oxygen content may also affect the intergrowth of RP phases\cite{Li2026bulk,Liu2023e,Zhang2020h}. Therefore, precisely controlling the oxygen content, identifying the true RP phases, and elucidating their corresponding pressure-induced superconducting properties are essential for understanding the underlying mechanisms.

Here, we have synthesized a series of high-quality La$_3$Ni$_2$O$_{7+\delta}$ ($-0.34 \leq \delta \leq 0.08$) polycrystalline samples with systematically controlled annealing conditions. Ambient-pressure Ni K-edge X-ray absorption fine structure (XAFS) measurements reveal that the oxygen content influences the tilting of the NiO$_6$ octahedra: as the oxygen content increases, the Ni-O-Ni bond angle along the $c$ axis decreases. Using neutron powder diffraction (NPD), SXRD, and STEM, we further demonstrate that oxygen content influences the intergrowth of RP phases. In the La$_3$Ni$_2$O$_{6.86}$ sample, the 1212 phase emerges alongside the bilayer phase, whereas the La$_3$Ni$_2$O$_{6.95}$ sample exhibits a nearly pure bilayer structure. For samples with higher oxygen content (La$_3$Ni$_2$O$_{7+\delta}$, -0.01 $\le$ $\delta$ $\le$ 0.08) obtained through oxygen annealing, the proportion of trilayer intergrowths increases significantly. Combined with high-pressure transport measurements, we observe superconducting transitions with distinct $T_c$s, which can be attributed to the bilayer phase, the 1212 phase, and trilayer inclusions. Furthermore, Ginzburg-Landau fitting analysis revealed that oxygen content directly modulates the $H_{c2}$ of the bilayer superconductivity, where a stoichiometric bilayer phase corresponds to a higher $H_{c2}$. The identification of superconductivity in distinct RP phases with varying oxygen contents provides new insights into the high-temperature superconducting mechanism of La$_3$Ni$_2$O$_{7}$.

\begin{figure*}[t]
	\includegraphics[width=\linewidth]{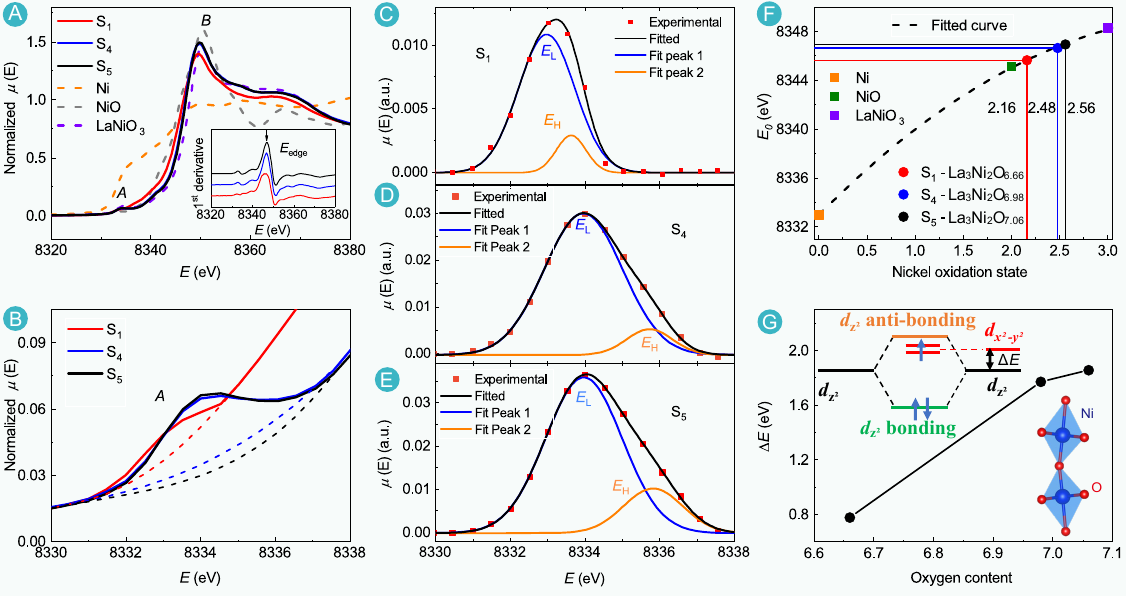}
	\caption{\label{fig1} Ni $K$-edge XAFS spectra (A) Experimental XAFS spectra of the reference samples Ni, NiO, LaNiO$_3$, and the experimental samples La$_3$Ni$_2$O$_{7+\delta}$ (S$_1$, S$_4$, and S$_5$). The alphabets $A$ and $B$ indicate the main features. Inset illustrates the first derivative function of the absorption coefficient $\mu(E)$ near the main absorption peak $B$. (B) Schematic diagram of background fitting for pre-edge peak $A$. (C-E) show the double Gaussian peak fitting of the pre-peak for experimental S$_1$, S$_4$ and S$_5$, respectively. $E_L$ and $E_H$ for the lower and higher energy peaks. (F) The valence information of the tested samples was determined by fitting with reference to the known valence states and absorption edge positions of reference samples. (G) The relationship between CFE and oxygen content in La$_3$Ni$_2$O$_{7+\delta}$. The top-left inset shows a schematic diagram of the degeneracy lifting in the energy levels of Ni's 3$d$ orbitals, while the bottom-right inset illustrates the nickel-oxygen octahedron in La$_3$Ni$_2$O$_{7+\delta}$.
	}
\end{figure*}

\textbf{MATERIALS AND METHODS }

\textbf{Material growth:}\ Six polycrystalline La$_3$Ni$_2$O$_{7+\delta}$ samples with varying oxygen content were synthesized and designated S$_1$ through S$_6$, ordered by their final oxygen stoichiometry. The starting materials were La$_2$O$_3$ (pre-sintered at 1000 $^\circ$C to remove moisture), NiO, and Ni(NO$_3$)$_2$$\cdot$6H$_2$O. S$_2$ was prepared by solid-state reaction. La$_2$O$_3$ and NiO were mixed in stoichiometric proportions, thoroughly grounded, and homogenized in an agate mortar. The mixture was placed in an alumina crucible and sintered at 1100 $^\circ$C for 3 days in air. The remaining samples (S$_1$, S$_3$-S$_6$) were synthesized via the sol-gel method. The as-grown parent S$_3$ was prepared as follows: stoichiometric amounts of La$_2$O$_3$ and Ni(NO$_3$)$_2$$\cdot$6H$_2$O were dissolved in dilute nitric acid. Citric acid was added, and the mixture was stirred until fully dissolved. The solution was heated to 100 $^\circ$C to evaporate water, forming a green gel. The gel was calcined at 800 $^\circ$C for 1 day to remove organic components, yielding a black powder. This powder was then pressed into pellets and sintered at 1100 $^\circ$C for 3 days in air. S$_4$ was obtained by annealing S$_3$ at 1100 $^\circ$C in flowing oxygen for 3 days. Annealing S$_4$ in H$_2$/N$_2$ (10$\%$ H$_2$) mixed gas at 400 $^\circ$C for 1 day yielded S$_1$. Annealing S$_4$ at 500 $^\circ$C under 100 bar oxygen pressure for 1 day and 5 days yielded S$_5$ and S$_6$, respectively.

\textbf{Neutron powder diffraction:}\ NPD measurements were performed using the time-of-flight General Purpose Powder Diffractometer (GPPD) located at the China Spallation Neutron Source (CSNS). The samples were loaded in 10-millimeter-diameter vanadium cans, and data were collected at room temperature. Data refinement was performed using the FULLPROF Suite\cite{Juan1993}.

\textbf{Thermogravimetric analysis:}\
TGA measurements were carried out on a Mettler$-$Toledo TGA/DSC 3$+$ instrument under a constant 80 mL/min flow of H$_2$/N$_2$ (4$\%$ H$_2$) mixed gas, with a heating rate of 5 $^\circ$C/min to 800 $^\circ$C. The accuracy of the TGA is 0.005\% of the measuring mass.

\textbf{High-pressure electronic transport:}\ High-pressure electronic transport measurement was conducted by using a 300 $\mu$m culet diamond anvil cell (DAC). The BeCu gasket was pre-indented to a thickness of 40 $\mu$m, and a 150 $\mu$m diameter sample hole was drilled. Polycrystal samples were cut into 50$\times$50$\times$20 $\mu$m$^3$ before loading. Solid-state pressure-transmitting medium KBr was adopted. 

\textbf{Scanning transmission electron microscopy:}\ STEM experiments were conducted using a double aberration-corrected JEOL ARM200F microscope operated at 200 kV. 

\textbf{Synchrotron radiation X-ray diffraction:}\ The SXRD experiments were performed at BL44B2 of SPring-8 with the approval of the Japan Synchrotron Radiation Research Institute (JASRI).

\textbf{X-ray absorption fine structure:}\ The Ni K-edge (E$_{Ni}$ = 11215 eV)  XAFS spectra were collected at the BM08 XAFS\cite{Acapito2019the} beamline of the European Synchrotron Radiation Facility (ESRF), Grenoble, France, at room temperature in standard transmission geometry. The Ni K-edge XAFS signals from the pure Ni-metal and the NiO reference powder samples were also measured simultaneously for comparison. For each sample, two spectra were collected, checked for energy calibration, and averaged, yielding high-quality data with low statistical noise. The experimental XAFS data have been treated according to the standard procedures for background subtraction, normalization, and extraction of structural XAFS signals\cite{Saha2021col,Bandyopadhyay2017,Meneghini2012}. The experiment code is HC-5447\cite{Alessandro2024,Bandyopadhyay2027}.

\textbf{RESULTS AND DISCUSSION }

XAFS spectroscopy was employed to analyze the oxidation state of Ni ions and the distortion of the NiO$_6$ octahedra. Three La$_3$Ni$_2$O$_{7+\delta}$ samples (S$_1$, S$_4$, S$_5$) and three reference samples (Ni, NiO, and LaNiO$_3$\cite{Alessandro2024}) were measured. The XAFS spectra in Figure \ref{fig1}A reveal two discernible features: a pre-edge peak $A$ and a main absorption peak $B$. Their origins have been discussed in previous studies \cite{Li2024dis,Li2025or,Mijit2024,Cai2025}. The pre-edge peak $A$ is attributed to both a minor quadrupole (1$s$ to 3$d$) component and, predominantly, a dipole-allowed (1$s$ to 4$p$) component. This dominant dipole transition arises because strong Ni-O orbital hybridization enhances the $p$ orbital character. The shape and fine structure of this peak are correlated with the local coordination symmetry of the NiO$_6$ octahedra. The main absorption peak $B$ is dominated by the dipole-allowed 1$s$ to 4$p$ transition and reflects the valence state of Ni ions.

\begin{figure}[t]
	\includegraphics[width=0.5\linewidth]{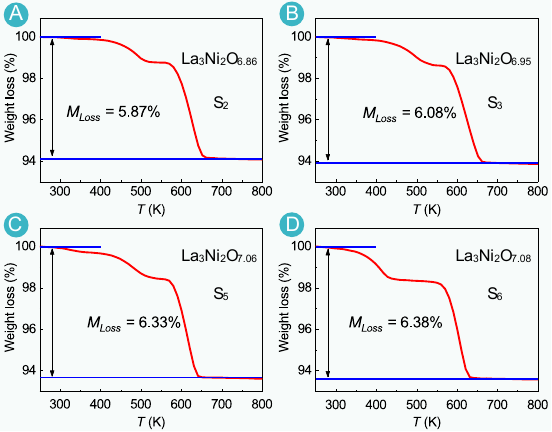}
	\caption{\label{fig2} Thermogravimetric analysis of La$_3$Ni$_2$O$_{7+\delta}$. TGA curves of the sample (A) S$_2$ (La$_3$Ni$_2$O$_{6.86}$), (B) S$_3$ (La$_3$Ni$_2$O$_{6.95}$), (C) S$_5$ (La$_3$Ni$_2$O$_{7.06}$), and (D) S$_6$ (La$_3$Ni$_2$O$_{7.08}$).}
\end{figure}

\begin{table}[h]
	\caption{A list of the sample number and corresponding measured oxygen contents for different methods. AVG. is an average of the oxygen content from the two methods.}
	\vspace{3mm}
	\begin{tabular}{ccccccc}
		\hline
		Method      &S$_1$&S$_2$&S$_3$&S$_4$&S$_5$&S$_6$
		\\ \hline
		XAFS   & 6.66 & -- &-- &6.98 &7.06 &-- \\
		
		TGA & --	&6.86 &6.95 &7.00 &7.06 &7.08\\
		
		AVG. & 6.66 &6.86 &6.95 &6.99 &7.06 &7.08 \\
		\hline \hline
	\end{tabular}
	\label{Table1}
\end{table}

\begin{figure*}[t]
	\includegraphics[width=\linewidth]{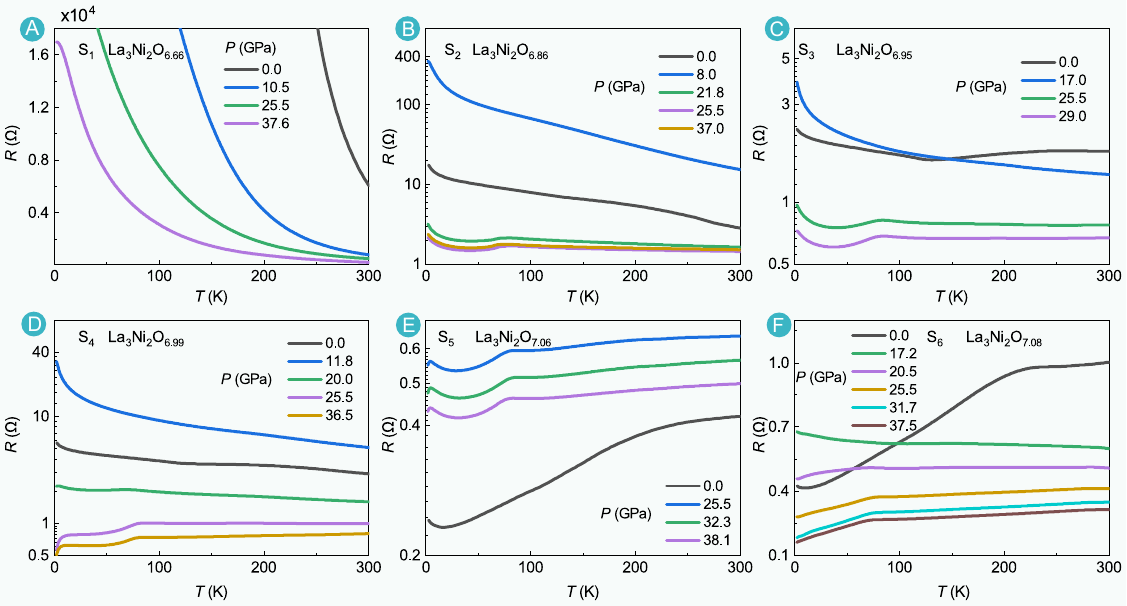}
	\caption{\label{fig3} Temperature-dependent resistance of La$_3$Ni$_2$O$_{7+\delta}$ samples under various pressures. The oxygen content of the sample is jointly determined based on XAFS and TGA results. (A) S$_1$ (La$_3$Ni$_2$O$_{6.66}$) exhibits insulating behavior up to 37.6 GPa. (B-F) S$_2$-S$_6$ (La$_3$Ni$_2$O$_{6.86}$-La$_3$Ni$_2$O$_{7.08}$) show pressure-induced superconductivity above 25 GPa with $T_c$ near 80 K.}
\end{figure*}

After subtracting a fitted background, Gaussian function fitting was applied to peak $A$ to extract the lower-energy peak ($E_L$) and higher-energy peak ($E_H$), as shown in Figures \ref{fig1}B-E. The crystal field splitting energy (CFE), defined as $\Delta E = E_H - E_L$, primarily corresponds to the energy splitting between the $d_{z^2}$ and $d_{x^2-y^2}$ orbitals\cite{Sakakibara2024p} and is closely related to the degree of NiO$_6$ octahedral distortion: a larger CFE indicates more pronounced distortion. The fitting results presented in Figure \ref{fig1}G reveal that in La$_3$Ni$_2$O$_{7+\delta}$, the CFE increases with oxygen content, implying larger tilting angle between two NiO$_6$ octahedra. This is consistent with previous findings that samples with more oxygen vacancies tend toward a tetragonal structure, whereas stoichiometric La$_3$Ni$_2$O$_7$ exhibits a Ni-O-Ni bond angle of 168$^\circ$, as illustrated in the bottom-right inset of Figure \ref{fig1}G.

\begin{figure}[t]
	\includegraphics[width=0.5\linewidth]{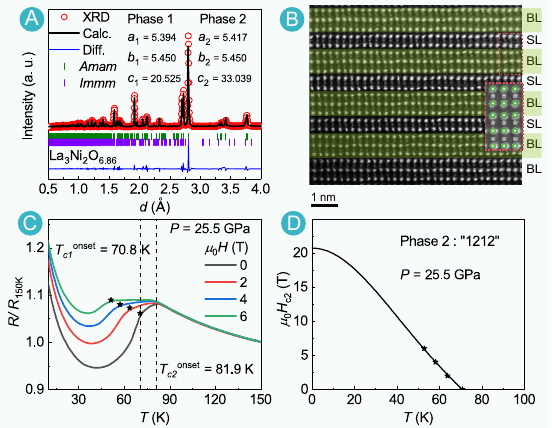}
	\caption{\label{fig4} Structure and high-pressure transport measurements of S$_2$ (La$_3$Ni$_2$O$_{6.86}$). (A) SR-XRD characterization of La$_3$Ni$_2$O$_{6.86}$. Phase 1 corresponds to a bilayer structure with the $Amam$ space group, while Phase 2 belongs to the $Immm$ space group and represents a 1212 phase. Phase 1: 76.1$\%$, Phase 2: 23.9$\%$. (B) STEM characterization of La$_3$Ni$_2$O$_{6.86}$. Green atoms represent La atoms and gray atoms represent Ni atoms in the red dashed box. SL and BL represent the intergrowth phases of La$_{n+1}$Ni$_n$O$_{3n+1}$ for $n=1, 2$. (C) High-pressure transport measurements at 25.5 GPa. The data reveal two distinct superconducting transitions. The transition temperature marked with an asterisk is associated with the 1212 structure. (D) Ginzburg-Landau fitting of the $H_{c2}$ for the superconducting transition associated with the 1212 structure, under applied magnetic fields.}
\end{figure}

The energy position of the main edge $B$ was used to determine the valence state of Ni ions. By using the first derivative method (inset of Figure \ref{fig1}A), the absorption edge energies were obtained for S$_1$, S$_4$, S$_5$, and the reference samples (Ni, NiO, and LaNiO$_3$). A calibration curve relating Ni valence to absorption edge energy was established (Figure \ref{fig1}F). From this calibration, the Ni valence states were determined to be $+2.16$, $+2.48$, and $+2.56$ for S$_1$, S$_4$, and S$_5$, respectively. The corresponding oxygen contents were calculated as 6.66, 6.98, and 7.06, respectively.

Thermogravimetric analysis (TGA) was performed on S$_2$--S$_6$ to determine their actual oxygen content. The results for S$_2$, S$_3$, S$_5$, and S$_6$ are presented in Table \ref{Table1}. S$_4$ and S$_5$ were measured using both TGA and XAFS. The valence state of Ni determined by XAFS and the oxygen content determined by TGA are consistent within the margin of error. Based on this combined analysis, the oxygen stoichiometries for S$_1$ through S$_6$ were determined to be 6.66, 6.86, 6.95, 6.99, 7.06, and 7.08, as listed in Table \ref{Table1}. We do not report individual errors, as the most significant uncertainty arises from the presence of intergrowth phases. Furthermore, we emphasize that the oxygen content determined by either XAFS or TGA represents a bulk average of the sample and does not correspond to the bilayer phase alone.

\begin{figure}[t]
	\includegraphics[width=\linewidth]{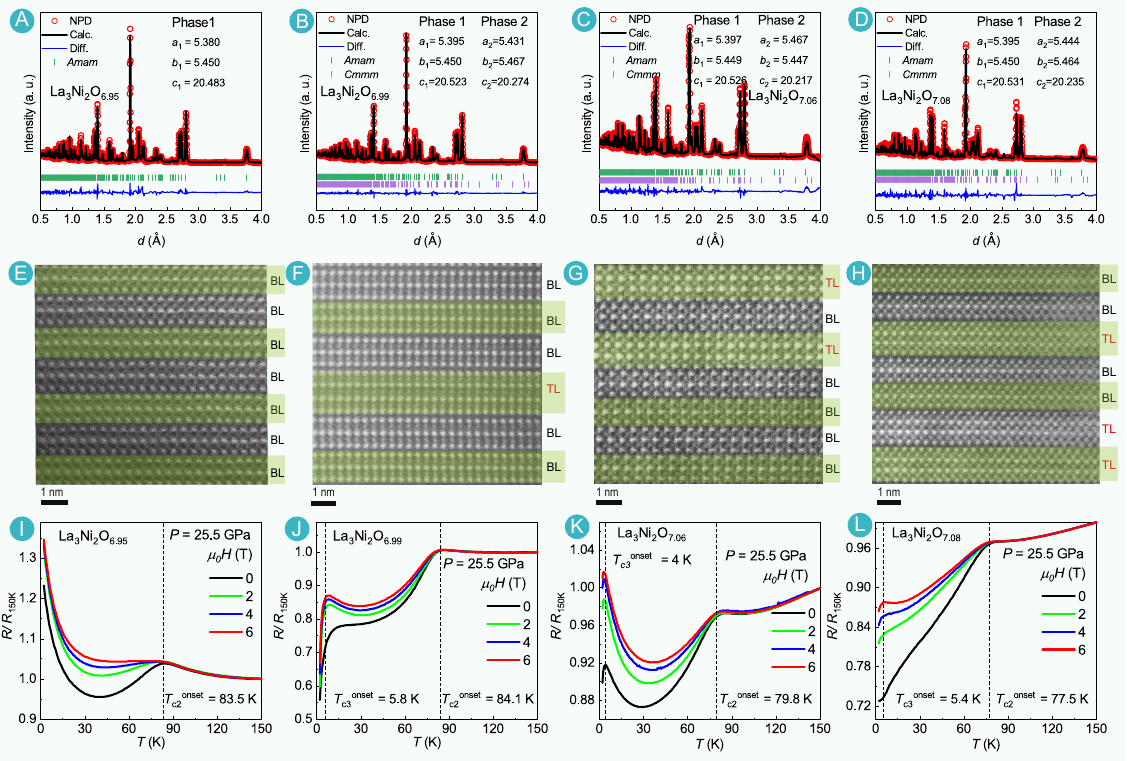}
	\caption{\label{fig5} Structural characterization and high-pressure transport measurements of S$_3$-S$_6$ (La$_3$Ni$_2$O$_{6.95}$-La$_3$Ni$_2$O$_{7.08}$). (A-D) NPD measurements of S$_3$ (La$_3$Ni$_2$O$_{6.95}$), S$_4$ (La$_3$Ni$_2$O$_{6.99}$), S$_5$ (La$_3$Ni$_2$O$_{7.06}$) and S$_6$ (La$_3$Ni$_2$O$_{7.08}$). Phase 1 corresponds to a bilayer structure with the $Amam$ space group, while Phase 2 belongs to the $Cmmm$  phase to mimic the randomly distributed trilayer intergrowths. S$_3$ (phase 1: 100$\%$), S$_4$ (phase 1: 88.2$\%$, phase 2: 11.8$\%$), S$_5$ (phase 1: 50.8$\%$, phase 2: 49.2$\%$) and S$_6$ (phase 1: 44.9$\%$, phase 2: 55.1$\%$). (E-H) Corresponding STEM characterization of samples matching the neutron diffraction patterns. TL represents the trilayer NiO$_6$. (I-L) Corresponding high-pressure transport measurements with various applied DC magnetic fields at 25.5 GPa of samples matching the neutron diffraction patterns. The high-pressure transport data of La$_3$Ni$_2$O$_{7+\delta}$ ($-0.01 \le \delta \le 0.08$) reveal two distinct superconducting transitions.}
\end{figure}

\begin{figure}[t]
	\includegraphics[width=0.5\linewidth]{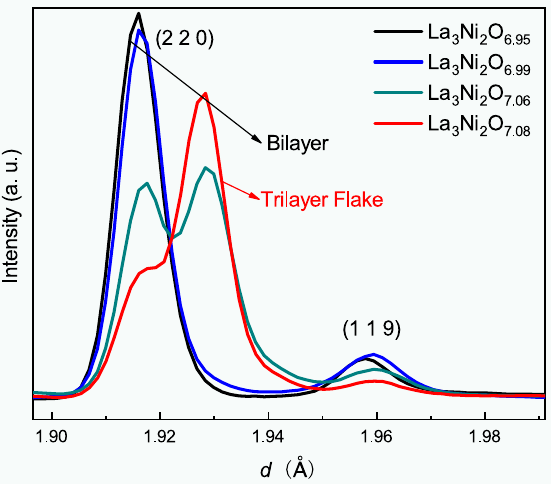}
	\caption{\label{fig6} Evolution of neutron diffraction peaks with oxygen content. The (2 2 0) diffraction develops a shoulder corresponding to trilayer intergrowths as oxygen content increases. The (1 1 9) peak shifts to higher $d$ with increasing trilayer fraction.The black line represents S$_3$ (La$_3$Ni$_2$O$_{6.95}$), the blue line represents S$_4$ (La$_3$Ni$_2$O$_{6.99}$), the green line represents S$_5$ (La$_3$Ni$_2$O$_{7.06}$), and the red line represents S$_6$ (La$_3$Ni$_2$O$_{7.08}$).}
\end{figure}

\begin{figure*}[t]
	\includegraphics[width=\linewidth]{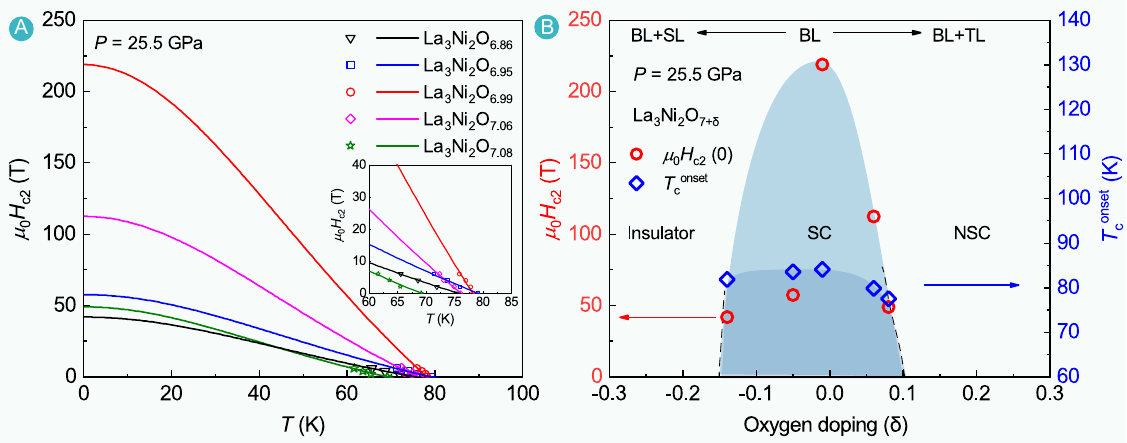}
	\caption{\label{fig7} Upper critical fields and phase diagram.
		(A) Ginzburg-Landau fitting of the upper critical fields for the bilayer superconducting phase in S$_2$-S$_6$ (La$_3$Ni$_2$O$_{6.86}$-La$_3$Ni$_2$O$_{7.08}$) at 25.5 GPa. (B) The phase diagram depicting the regulation of RP structure and bilayer superconductivity by oxygen content. 
		The middle region indicates the range of the pure BL phase. In La$_3$Ni$_2$O$_{7+\delta}$, no superconductivity is observed under pressure when ${\delta}$ = -0.15 and 0.23.
		The blue shaded region within the dashed lines indicates the superconducting (SC) region, while the white region marks the nonsuperconducting (NSC) region. The red circle data points obtained from (A), and the blue diamond data points obtained from Figure \ref{fig3} and \ref{fig5}.
	}
\end{figure*}

Systematic high-pressure electronic transport measurements were performed on all six compositions. S$_1$, which has the lowest oxygen content, exhibits insulating behavior across the entire pressure range measured (Figure \ref{fig3}). Theoretical studies suggest that La$_3$Ni$_2$O$_{6.5}$ exhibits Mott insulator characteristics\cite{Zhang2024ele1,Wang2025self}, consistent with the experimental results. As the oxygen content increases, the samples become progressively more metallic at ambient pressure. Under applied pressures above 25 GPa, the resistance of S$_2$-S$_6$ shows a clear drop near 80 K, providing evidence of superconductivity originating from the bilayer structure of La$_3$Ni$_2$O$_{7+\delta}$. To further characterize the superconducting properties of the bilayer phase, we conducted detailed structural and electronic transport measurements on the pressurized superconducting compounds.

Figure \ref{fig4} presents the structural and transport characterization of S$_2$. As shown in Figure \ref{fig4}A, SXRD Rietveld refinement indicates that while the main phase in La$_3$Ni$_2$O$_{6.86}$ is the bilayer structure with space group $Amam$, the sample also contains a secondary phase identified as the 1212 structure with space group $Immm$.\cite{Li2024de} To verify the presence of the 1212 phase at the microscopic scale, we performed STEM measurements, as shown in Figure \ref{fig4}B \& S1. The 1212 phase is clearly resolved in S$_2$.
It is known that the bilayer and 1212 structures exhibit distinct $T_c$ under high pressure. Indeed, our high-pressure electrical transport measurements at 25.5 GPa reveal two superconducting transitions, with onset temperatures $T_{c1}^{\mathrm{onset}} = 70.8$ K and $T_{c2}^{\mathrm{onset}} = 81.9$ K (Figure \ref{fig4}C). The lower-temperature transition ($T_{c1}^{\mathrm{onset}}$) is significantly suppressed under an applied magnetic field, analogous to the superconducting behavior previously reported for the 1212 phase. Fitting the $H_{c2}$ using the Ginzburg-Landau formula yields $\mu_0H_{c2}(0) = 20.7$ T for the 1212 phase (Figure \ref{fig4}D).

To investigate the relationship among composition, structure, and superconductivity in RP nickelates, we performed comprehensive NPD and STEM measurements on S$_3$-S$_6$, as shown in Figures \ref{fig5}A-D. S$_3$, synthesized directly in air via the sol-gel method, has a stoichiometry of La$_3$Ni$_2$O$_{6.95}$. The NPD refinement in Figure \ref{fig5}A reveals a single bilayer phase. STEM imaging further confirms the bilayer arrangement at atomic resolution over tens of nanometers (Supplementary Figure S2A). High-pressure transport measurements at 25.5 GPa under a magnetic field reveal a superconducting transition at 83.5 K, consistent with the $T_c$ reported in pressurized single crystals.
For S$_4$-S$_6$, both NPD (Figures \ref{fig5}B-D) and STEM (Figures \ref{fig5}F-H \& S2B for S$_5$) results show that the proportion of trilayer intergrowths increases with oxygen content. The NPD data can be well fitted by a two-phase model comprising a bilayer structure (space group $Amam$) and a monolayer-trilayer structure (space group $Cmmm$). Notably, phase 2, representing the trilayer intergrowths, appears progressively and is randomly distributed within the bilayer matrix. Within the bilayer structure, the lattice parameter $c$ increases with oxygen content, from 20.483 to 20.531 \AA.
Transport measurements at 25.5 GPa reveal two distinct superconducting transitions (Figures \ref{fig5}J-L): $T_{c2}^{\mathrm{onset}}$, originating from the bilayer phase, and $T_{c3}^{\mathrm{onset}}$, associated with the trilayer intergrowths. The transition temperatures of these trilayer intergrowths range from 4 to 6 K, which is close to the $T_c$ of 3.6 K reported for the pure monolayer-trilayer La$_3$Ni$_2$O$_7$\cite{Huang2025s}.

We examined the effect of oxygen content on structural evolution in the NPD patterns. Figure \ref{fig6} shows the evolution of the (2 2 0) and (1 1 9) diffraction peaks for S$_3$-S$_6$. As the oxygen content and the proportion of trilayer intergrowths increase, a new diffraction emerges near the (2 2 0) peak of the bilayer structure. This new feature can be attributed to the (2 2 0) diffraction of trilayer intergrowths, consistent with their larger in-plane lattice constant compared to the bilayer phase. The evolution of the NPD peaks is in good agreement with STEM observations, where the relative abundance of trilayer intergrowths becomes comparable to that of the bilayer matrix. Correspondingly, the broadening of the (1 1 9) peak intensifies as the volume fraction of trilayer intergrowths increases.

To further investigate the effect of oxygen content and intergrowth phases on the superconductivity of the bilayer structure, we present the $H_{c2}$ of the bilayer superconductivity as a function of the averaged oxygen content of the bulk sample in Figure \ref{fig7}. The Ginzburg-Landau fittings are based on electronic transport measurements performed at 25.5 GPa, shown in Figures \ref{fig4}C and \ref{fig5}I-L. We adopted the 90\% resistance drop criterion for defining $H_{c2}$ to enable systematic comparison across samples, as the presence of an insulating background in some samples made alternative methods less reliable. The results reveal that for oxygen content below 7, $H_{c2}$ increases with increasing oxygen content. Although S$_3$ and S$_4$ have similar oxygen contents, the lower $H_{c2}$ in S$_3$ may be attributed to the presence of oxygen vacancies. In contrast, for oxygen content above 7, $H_{c2}$ decreases. While the superconducting transition temperature $T_c$ exhibits only modest variation, the $H_{c2}$ changes significantly. Based on previous reports, pressure-induced superconductivity in La$_3$Ni$_2$O$_{7+\delta}$ is absent for $\delta \leq -0.15$ and $\delta \geq 0.23$ \cite{Ueki2025, Dong2025i}.

As the oxygen content increases, two concurrent effects influence the bilayer phase. First, the number of apical oxygen vacancies decreases, as reflected by the increase in the out-of-plane lattice constant $c$. Second, the density of trilayer intergrowths increases. For $\delta < 0$, the suppression of $H_{c2}$ is attributed to apical oxygen vacancies and the presence of single-layer NiO$_6$ octahedra intergrowths. For $\delta > 0$, the reduction in $H_{c2}$ is associated with the increasing proportion of trilayer intergrowths. It is important to note that the reported oxygen content represents an average over the entire sample. If one considers the presence of intergrowth phases, the actual oxygen doping range over which the bilayer phase remains superconducting would be narrower than indicated in Figure \ref{fig7}B.

\textbf{CONCLUSION}

\vspace{1em}

In summary, we have demonstrated regulation of the oxygen content in La$_3$Ni$_2$O$_{7+\delta}$ polycrystalline samples, enabling systematic tuning of both the Ruddlesden-Popper intergrowth structures and their superconducting properties. Through comprehensive X-ray absorption fine structure, neutron powder diffraction, and scanning transmission electron microscopy measurements, we established that oxygen content governs two critical aspects: the distortion of NiO$_6$ octahedra and the formation of intergrowth phases. Low oxygen content promotes the emergence of monolayer-bilayer intergrowths, while high oxygen content favors trilayer inclusions within the bilayer matrix.

High-pressure transport measurements reveal distinct superconducting signatures associated with different RP phases. The bilayer phase exhibits superconductivity near 80 K, the 1212 phase shows a transition around 70 K, and trilayer intergrowths display superconductivity at 4--6 K, consistent with the recently discovered pure monolayer-trilayer phase. Most importantly, we find that the $H_{c2}$ of the bilayer superconductivity is strongly modulated by oxygen content, peaking near the stoichiometric composition ($\delta \approx 0$) and decreasing on both underdoped and overdoped sides due to the presence of intergrowth defects.

These findings establish a comprehensive phase diagram linking oxygen stoichiometry, structural intergrowths, and superconducting properties in La$_3$Ni$_2$O$_{7+\delta}$. Our work not only advances the synthesis protocols for phase-pure RP nickelates but also provides crucial insights into the role of interlayer coupling and apical oxygen in mediating high-temperature superconductivity. The demonstrated sensitivity of superconductivity to local structural variations highlights the importance of precise stoichiometric control in the search for and optimization of new nickelate superconductors.

\vspace{1em}

\textbf{REFERENCES}

\begingroup
\renewcommand{\section}[2]{}

\vspace{1em}
\textbf{FUNDING AND ACKNOWLEDGMENTS}\\
This work was supported by the National Natural Science Foundation of China (No. 12425404, U25A20193, 12474137, 12494591, 92565303, 52273227, 112504170), the National Key Research and Development Program of China (Grants No. 2023YFA1406500, 2023YFA1406000), the Fundamental and Interdisciplinary Disciplines Breakthrough Plan of the Ministry of Education of China (JYB2025XDXM403),the CAS Superconducting Research Project (Grants No. SCZX-0101), the Guangdong Basic and Applied Basic Research Funds (Grants No. 2024B1515020040, 2025B1515020008, 2024A1515030030), the Guangdong Major Project of Basic Research (2025B0303000004), the Guangzhou Basic and Applied Basic Research Funds (Grant No. 2024A04J6417), Shenzhen Science and Technology Program (Grants No. RCYX20231211090245050), Guangdong Provincial Key Laboratory of Magnetoelectric Physics and Devices (Grant No. 2022B1212010008), the Guangdong Provincial Quantum Science Strategic Initiative (Grant No. GDZX2501006), and Research Center for Magnetoelectric Physics of Guangdong Province (Grants 2024B0303390001). A.B. and D.T.A. thank the EPSRC UK for the funding (Grant No. EP/W00562X/1). A.B. further acknowledges Lalit. Narayan Mithila University for giving financial support. D.T.A. thanks the CAS for PiFi funding. We gratefully acknowledge the neutron beam time provided by the General Purpose Powder Diffractometer (GPPD) at the China Spallation Neutron Source (CSNS) and the scientific guidance on neutron diffraction from the GPPD staff. We acknowledge the beam time for X-ray Absorption Fine Structure (XAFS) spectroscopy provided by the BM08 beamline (The LISA Beamline) at the European Synchrotron Radiation Facility (ESRF) and thank Dr. Francesco d'Acapito and the BM08 staff for their assistance. We also acknowledge the X-ray Diffraction (XRD) beam time at the SPring-8 facility provided by the Japan Synchrotron Radiation Research Institute (JASRI) (Proposal Nos. 2025A1497).

\vspace{1em}
\textbf{AUTHOR CONTRIBUTIONS}\\
Peiyue Ma: Investigation, Writing - original draft, Data curation, Conceptualization, Visualization, Formal analysis. 	
Jingyuan Li: Conceptualization, Investigation, Writing - original draft, Data curation, Visualization, Formal analysis. 
Xing Huang: Conceptualization, Data curation, Methodology, Visualization, Writing - original draft. 
Yixing Zhao: Methodology, Data curation, Formal analysis. Yifeng Han: Methodology, Formal analysis.
Mengwu Huo:	Methodology.
Deyuan Hu: Methodology, Data curation.
Chaoxin Huang: Methodology, Investigation.
Hengyuan Zhang: Methodology, Investigation.
Sihao Deng: Methodology, Formal analysis, Data curation.
Lunhua He: 	Methodology, Visualization, Data curation.
Juan Rodriguez-Carvajal: Methodology, Formal analysis.
Abhisek Bandyopadhyay: Methodology, Formal analysis, Data curation.
Alessandro Puri: Methodology, Visualization, Data curation, Investigation.
Devashibhai Adroja: Investigation, Methodology, Visualization, Formal analysis, Data curation.
Xiang Chen:	Methodology.
Tao Xie: Methodology.
Zhen Chen: Funding acquisition, Methodology, Visualization, Data curation.
Hualei Sun: Supervision, Conceptualization, Funding acquisition, Writing - original draft, Writing - review and editing, Visualization, Methodology, Formal analysis, Data curation, Resources.
Meng Wang: Supervision, Project administration, Conceptualization, Funding acquisition, Writing - original draft, Writing - review and editing, Resources, Methodology, Investigation, Visualization, Formal analysis, Data curation, Validation.
All authors contributed to the manuscript and approved the final version.

\vspace{1em}
\textbf{DECLARATION OF INTERESTS}\\
The authors declare no conflicts of interest.\\

\vspace{1em}
\textbf{DATA AND CODE AVAILABILITY} \\
Data are available from the corresponding author upon reasonable request.

\newpage

\end{document}